\documentstyle[11pt,newpasp,twoside,epsf]{article}
\def\edcomment#1{\iffalse\marginpar{\raggedright\sl#1\/}\else\relax\fi}
\reversemarginpar

\begin{document}
\title{Constraints on Quasar Continuum, BELR, and BALR Physics from SDSS Composite Spectra}
\author{Gordon T. Richards}
\vspace{-0.05cm}
\affil{Princeton University Observatory, Princeton, NJ 08544-1001}
\vspace{-0.05cm}
\author{Patrick B. Hall}
\vspace{-0.05cm}
\affil{Princeton University Observatory, Princeton, NJ 08544-1001 and
Departamento de Astronom\'{\i}a y Astrof\'{\i}sica, Facultad de F\'{\i}sica,
Pontificia Universidad Cat\'{o}lica de Chile, Casilla 306, Santiago 22, Chile}
\vspace{-0.05cm}
\author{Timothy A. Reichard}
\vspace{-0.05cm}
\affil{Johns Hopkins University, Department of Physics and Astronomy, 3400 N. Charles St., Baltimore, MD 21218}
\vspace{-0.05cm}
\author{Daniel E. Vanden Berk}
\vspace{-0.05cm}
\affil{Univ. of Pittsburgh, Dept. of Physics and Astronomy, 3941 O'Hara St., Pittsburgh, PA 15260} 
\vspace{-0.05cm}
\author{Donald P. Schneider}
\vspace{-0.05cm}
\affil{The Pennsylvania State University, Department of Astronomy and Astrophysics, 525 Davey Lab, University Park, PA 16802}
\vspace{-0.05cm}
\author{Michael A. Strauss}
\vspace{-0.05cm}
\affil{Princeton University Observatory, Princeton, NJ 08544-1001}
\vspace{-0.05cm}

\begin{abstract}
We review recent results on quasars from the SDSS as they relate to our
understanding of the UV/optical continuum, the broad emission line
region, and the broad absorption line region.  The ensemble average
colors of large numbers of quasars promise to provide constraints on the
optical/UV continuum emission mechanism.  High-ionization
emission-line blueshifts and emission line properties as a function of
optical/UV spectral index trace the structure of the broad emission
line region.  Statistical analysis of the broad absorption line quasar
population suggests that they are more ubiquitous than one might
otherwise think and are not likely to represent a completely distinct
population of quasars, but that the BAL trough properties are a
function of the underlying optical continuum and emission properties of the
quasar.

\end{abstract}

\vspace{-0.3cm}

\section{The Optical/UV Continuum}

In the rest-frame optical and UV, after accounting for emission lines,
quasars are reasonably well-described by a power-law continuum with a
typical spectral index of $\alpha_\nu\sim-0.3$--$0.5$ (e.g., Francis et
al.\ 1991; Vanden Berk et al.\ 2001).

Using data from Sloan Digital Sky Survey (SDSS; York et al.\ 2003)
quasars (Richards et al.\ 2002a), we find that the spread in the color
distribution as measured by SDSS quasars is roughly
$\Delta\alpha_\nu=\pm0.25$ ($1\sigma$) and is formally resolved
(Richards et al.\ 2003).  That is, the width of the distribution is
much broader than the errors.  Detailed analysis of the contributions
of dust extinction (which is the most likely cause of the red tail in
Figure~1; Hall et al.\ 2004a, this volume) and variability
(Wilhite et al.\ 2004, this volume) are needed in order to accurately
describe the intrinsic continua of quasars; however, the raw data
already provide constraints for accretion disk models.  We can confirm
that the average color is significantly redder than is expected from a
simplistic accretion disk modeled as a sum of blackbodies
($\alpha_\nu=1/3$; e.g., Shakura \& Sunyaev 1973; Hubeny et al.\
2000), but the bluest quasars can be even bluer than $\alpha_\nu=1/3$.
Comparisons of the distribution of colors to accretion disk models
will provide an important constraint on those models (e.g., Blaes 2004, this volume).

\begin{figure}
\plotfiddle{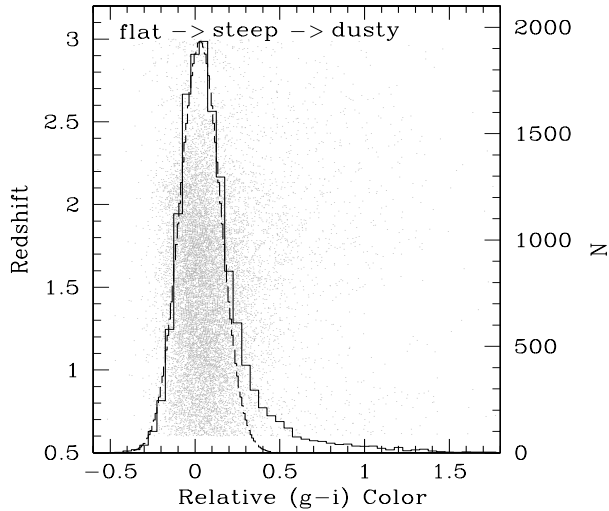}{2.4in}{0.0}{40}{35}{-110}{-55}
\caption{Distribution of relative $g-i$ colors (observed colors
corrected by the median color as a function of redshift) for the SDSS
DR1 quasar sample. Note the red tail that extends beyond an otherwise
Gaussian distribution. (Adapted from Richards et al.\ 2003.)
\label{fig:fig1}}
\end{figure}

\section{Broad Emission Line Region}

The SDSS data can help improve our understanding of the broad emission
line region (BELR) by studying the changes in the strengths and
profiles of the broad emission lines as a function of their rest-frame
optical/UV colors and as a function of the well-known blueshift of
high-ionization emission lines (especially C~IV) with respect to
low-ionization and forbidden, narrow emission lines (Gaskell 1982;
Richards et al.\ 2002b).

The SDSS data reveal that that blueshift of C~IV may be caused by a
lack of flux in the red wing rather than by a bulk blueward shift of
the line since the blueshifted lines are also systematically weaker
--- perhaps due to an orientation and/or radiative transfer effect in
a disk-wind (e.g., Chiang \& Murray 1996).  Although the composites in
Figure~2 are normalized by redshift and luminosity, the fact that the
blueshifted composites have weaker C~IV emission suggests that the
emission line blueshifts and the Baldwin (1977) Effect have the same
physical basis.   See Francis \& Koratkar (1995) for confirmation
of this relationship in another sample of quasars.  Not correcting
redshifts for these blueshifts may be the reason that SPCA analyses
(Shang \& Wills 2004, this volume; Yip et al.\ 2004, this volume) find
a Baldwin Effect in the apparent line cores rather than the red wing
of the line; see Vanden Berk et al.\ (2004, this volume) for
supporting evidence.  Furthermore, how well one can measure the mass
of a quasar from its C~IV profile clearly depends on the exact nature
of these blueshifts (c.f., Vestergaard 2004a, this volume; Warner,
Hamann, \& Dietrich 2004, this volume).

\begin{figure}[t]
\plotfiddle{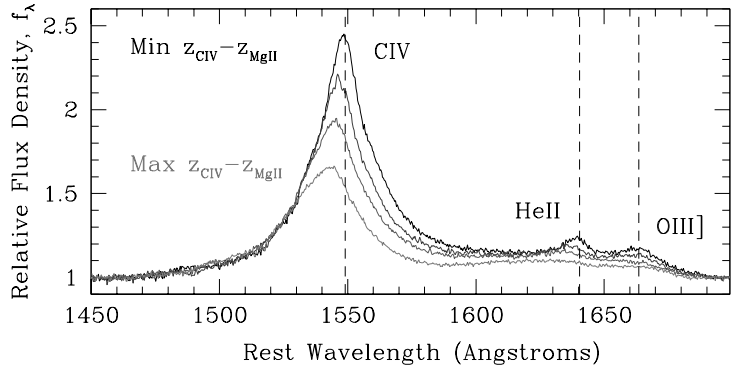}{1.6in}{0.0}{50}{50}{-145}{-85}
\caption{Composite spectra as a function of the redshift difference
between CIV and MgII. (Adapted from Richards et al.\ 2002b.) \label{fig:fig2}}
\vspace{-0.2cm}
\end{figure}

\begin{figure}[h!]
\vspace{0.1cm}
\plotfiddle{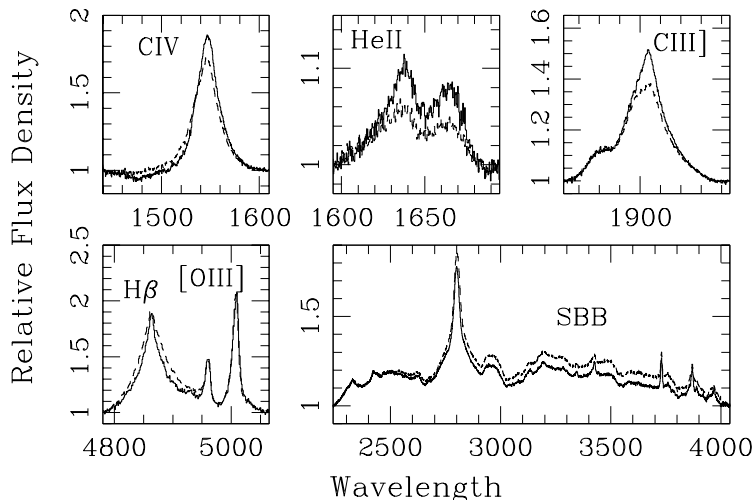}{2.1in}{0.0}{50}{50}{-145}{-85}
\caption{Selected emission line regions of composite spectra of the
reddest (solid line) and bluest (dashed line) SDSS quasars [excluding
those that appear dust reddened]. (Adapted from Richards et al.\ 2003.)\label{fig:fig3}}
\end{figure}

We also find that quasar emission lines are a function of the
optical/UV continuum slope (i.e., color). Figure~3 shows major
emission line regions for composite spectra constructed from quasars
with bluer and redder than average relative $g-i$ colors
(Richards et al.\ 2003).  More work needs to be done to understand
this relationship, but it is clear that these differences are not
(solely) due to the redder quasars (those denoted `steep' in Fig.~1)
having more dust.

\section{Broad Absorption Line Region}

The SDSS quasar survey has already provided the largest sample to date
of broad absorption line quasars.  The most significant results are as
follows.  The balnicity index (Weymann et al.\ 1991) distribution
increases steeply with decreasing absorption strength (Tolea, Krolik,
\& Tsvetanov 2003; Reichard et al.\ 2003a), which suggests that the
true population of quasars with intrinsic absorption outflows is
larger than is generally thought and also that quasars with wide
high-velocity outflows may not be distinct from those with narrower
low-velocity outflows.  Analysis of the continuum and emission line
regions of SDSS BALQSOs by Reichard et al.\ (2003b) suggests that the
optically selected BALQSO sample is drawn from the same parent
population as the nonBALQSO sample, but it does appear that the
properties of BALQSOs (terminal velocity, ionization state, etc.) are
not independent of the intrinsic color and emission line properties.
Thus even though BALQSOs may not be a distinct population, certain
types of quasars may be more likely to host certain kinds of BALs.
For example, composites of intrinsically red and intrinsically blue
BALQSOs (after correction for dust reddening) appear to have somewhat
different broad absorption trough properties; see Figure~4.

\begin{figure}[t]
\plotfiddle{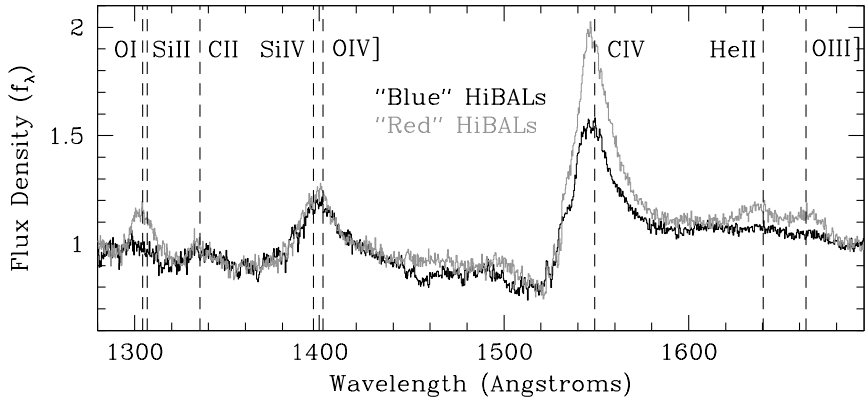}{1.9in}{0.0}{60}{60}{-180}{-100}
\caption{Composite spectra of HiBALs with intrinsically blue spectral
indices ({\em black line}) and intrinsically red spectral indices
({\em grey line}).  Each composite is the average of 41
quasars. (Adapted from Reichard et al.\ 2003.)}
\end{figure}

\section{Support for a Hybrid Model?}

We suggest that the combination of the results from SDSS BALQSOs above
and previous results (such as the difference between narrow absorption
in flat- and steep-spectrum radio-loud quasars) lends support to a
model where the disk-wind ranges from being nearly polar to nearly
equatorial (with BAL or BAL-like outflows existing in all objects);
see Figure~5.  Such a scenario might result from a hybrid model which
combines MHD and line-driven (LD) radiation pressure such as discussed
by both Proga (2003) and Everett (2003).  If this were the case, the
Elvis (2000) picture (with polar NAL absorption regions and more
equatorial BAL absorption regions; Figure~5, top) might be seen as
being the ensemble average picture, rather than the picture of an
individual quasar.  Elvis (2000) discusses similar scenarios in
relation to incorporating radio-loud quasars into the picture and how
luminosity might affect the opening angle of the wind.  Such a picture
would create two orientation effects (the opening angle of the wind
and the tilt of the disk) that may be difficult to disentangle;
however such a scenario would open up some freedom to explain all
known classes of AGN using a disk-wind model.

\begin{figure}[t]
\plotfiddle{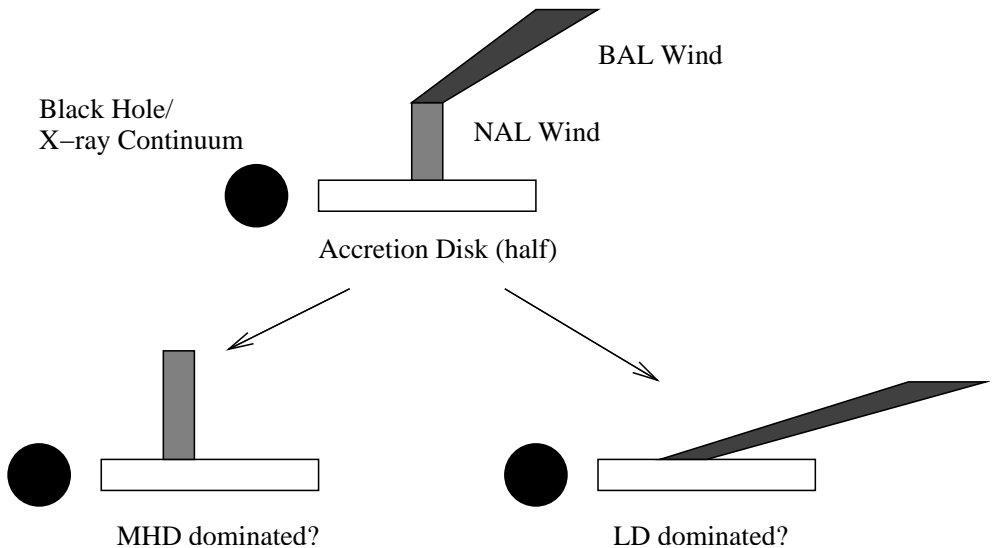}{2.8in}{0.0}{65}{65}{-180}{0}
\caption{Absorption line data suggests that quasars may all have
BAL-like outflows, but that their properties depend on the opening
angle of the wind which may range from equatorial to polar (perhaps
depending on the relative strength of the disk magnetic fields and
radiation pressure).}
\end{figure}

\acknowledgments
Funding for the creation and distribution of the SDSS Archive
(http://www.sdss.org/) has been provided by the Alfred P. Sloan
Foundation, the Participating Institutions, the National Aeronautics
and Space Administration, the National Science Foundation, the
U.S. Department of Energy, the Japanese Monbukagakusho, and the Max
Planck Society.

The SDSS is managed by the Astrophysical Research Consortium (ARC) for
the Participating Institutions. The Participating Institutions are The
University of Chicago, Fermilab, the Institute for Advanced Study, the
Japan Participation Group, The Johns Hopkins University, Los Alamos
National Laboratory, the Max-Planck-Institute for Astronomy (MPIA),
the Max-Planck-Institute for Astrophysics (MPA), New Mexico State
University, University of Pittsburgh, Princeton University, the United
States Naval Observatory, and the University of Washington.

\question{Antonucci} How much of the spread in colors do you think is
due to variability of individual objects as they vary in luminosity?

\answer{Richards} Variability certainly contributes significantly to
the spread, but it is not the dominant factor.  We estimate that it
contributes about $0.05$ (out of $\sim0.12$) to the $1\sigma$ spread
in color (or about $0.1$ of $0.25$ in $\Delta\alpha$).

\end{document}